\documentclass[fleqn,usenatbib]{mnras}

\usepackage{newtxtext,newtxmath}

\usepackage[T1]{fontenc}

\DeclareRobustCommand{\VAN}[3]{#2}
\let\VANthebibliography\thebibliography
\def\thebibliography{\DeclareRobustCommand{\VAN}[3]{##3}\VANthebibliography}

\usepackage{graphicx}	
\usepackage{amsmath}	

\title[Primordial nucleosynthesis constraints]{Primordial nucleosynthesis constraints on high-$z$ energy releases}

\author[G. De Zotti et al.]{Gianfranco De Zotti$^{1}$\thanks{E-mail: gianfranco.dezotti@inaf.it} and
Matteo Bonato,$^{2,1}$
\\
$^{1}$INAF--Osservatorio Astronomico di Padova, Vicolo dell'Osservatorio 5, I-35122 Padova, Italy\\
$^{2}$INAF--Istituto di Radioastronomia and Italian ALMA Regional Centre, Via Gobetti 101, I-40129, Bologna, Italy\\
}

\date{Accepted XXX. Received YYY; in original form ZZZ}

\pubyear{2015}

\begin{document}
\label{firstpage}
\pagerange{\pageref{firstpage}--\pageref{lastpage}}
\maketitle

\def\simlt{\mathrel{\rlap{\lower 3pt\hbox{$\sim$}}\raise 2.0pt\hbox{$<$}}}
\def\simgt{\mathrel{\rlap{\lower 3pt\hbox{$\sim$}} \raise 2.0pt\hbox{$>$}}}

\begin{abstract}

The cosmic microwave background (CMB) spectrum provides tight constraints on
the thermal history of the universe up to $z \sim 2\times 10^6$. At higher
redshifts thermalization processes become very efficient so that even large
energy releases do not leave visible imprints in the CMB spectrum. In this
paper we show that the consistency between the accurate determinations of the
specific entropy at primordial nucleosynthesis and at the electron-photon
decoupling implies that no more than 7.8\% of the present day CMB energy
density could have been released in the post-nucleosynthesis era. As pointed
out by previous studies, primordial nucleosynthesis complements model
independent constraints provided by the CMB spectrum, extending them by two
orders of magnitude in redshift.

\end{abstract}




\section{Introduction}

The spectrum of the cosmic microwave background (CMB) carries unique
information/constraints on the thermal history of the universe since energy
releases occurring over many redshift decades can leave their imprint on it
\citep{ZeldovichSunyaev1969, SunyaevZeldovich1970, IllarionovSunyaev1975,
DaneseDeZotti1977, Burigana1991b, Chluba2012, KhatriSunyaev2012,
ChlubaJeong2014, Tashiro2014, Chluba2016, DeZotti2016, Chluba2019}. However, at
very high redshifts such imprints are erased by thermalization effects due to
the combined action of photon emission processes and of Compton scattering.
Small distortions are completely thermalized at $z > \hbox{few}\times 10^6$
\citep{DaneseDeZotti1982, Burigana1991b, HuSilk1993a, Chluba2014a}. The
thermalization is less efficient for large distortions which can keep some
visibility up to redshifts higher by factors of several \citep{Chluba2020}.

At still higher redshifts, releases of very large amount of energy could have
occurred without leaving any visible track in the CMB spectrum. On the other
hand, the outcomes of primordial nucleosynthesis (or big-bang nucleosynthesis,
BBN) are sensitive to the radiation energy density. Indeed, just the
consideration of the production of light elements in the early universe led to
the prediction of the CMB \citep{Gamow1948, AlpherHerman1949}.

The present-day accurate determinations of cosmological parameters entail
strong constraints on the CMB energy density both at the BBN epoch and at
electron-photon decoupling, hence on the additional amount of energy that could
have been released after the BBN, at high enough redshifts to ensure a tight
coupling between electrons and radiation. Such constraints are quantified in
Sect.~\ref{sect:results}. Our conclusions are summarized in
Sect.~\ref{sect:conclusions}.

\begin{figure}
	\includegraphics[width=\columnwidth]{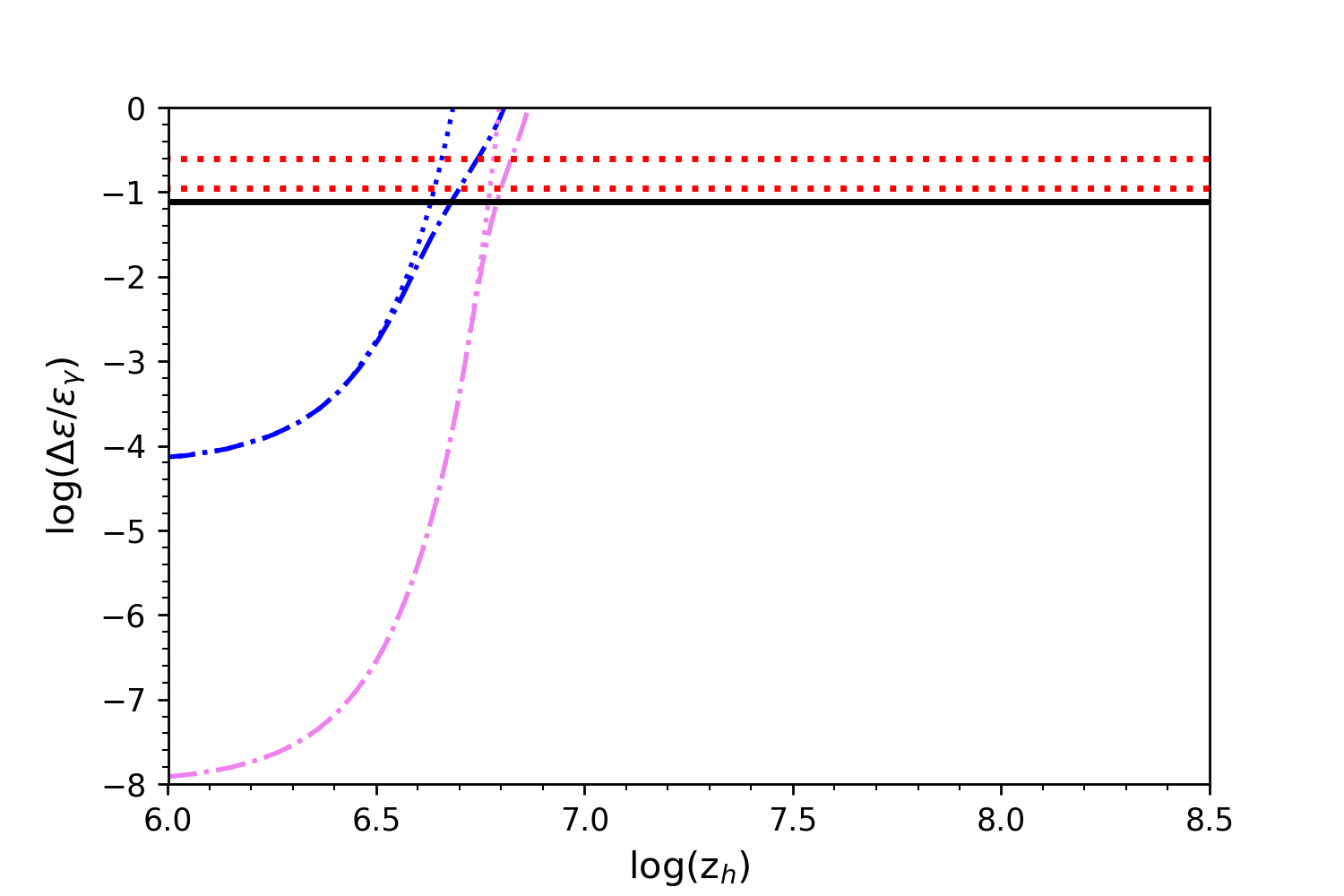}
    \caption{Upper limit (95\% confidence level) to the fractional amount of energy
    that could have been added to the CMB after the primordial nucleosynthesis epoch, derived in this paper (black horizontal line),
    compared with constraints derived by \citet{Chluba2020}. The red dotted horizontal lines show the limits
    obtained from the effective number of neutrino species, assuming that a
fraction $f_\nu$ of the total injected energy goes into neutrinos. The lower and upper lines refer to
$f_\nu=0$ and $f_\nu=0.5$, respectively; note that the value for $f_\nu=0$ has been corrected
as mentioned in the text. Constraints
    from CMB spectral measurements, as computed by \citet{Chluba2020}, are also shown.
    The upper (blue) and lower (pink) dot-dashed lines show the 95\% confidence upper limits
    implied by CMB spectral measurements from COBE/FIRAS ($\delta\epsilon/\epsilon_\gamma< 6\times 10^{-5}$)
    and from a future PIXIE-like experiment ($\delta\epsilon/\epsilon_\gamma< 10^{-8}$),
    respectively. The dotted lines deviating from the dot-dashed lines at their right end, again from \citet{Chluba2020},
    show the corresponding limits obtained under the small-distortion approximation.}
    \label{fig:constraints}
\end{figure}

\section{Constraints on the radiation energy density at
BBN}\label{sect:results}

The results of BBN calculations can be presented as a function of the
present-day dimensionless baryon-to-photon number density ratio $\eta =
n_b/n_\gamma$ \citep{Steigman2007, Cyburt2016, Fields2020}. The BBN redshift
($z_{\rm BBN}\sim 3\times 10^8$) is much higher than the CMB thermalization
redshift even in the case of strong deviations from equilibrium
\citep{Chluba2020}. Hence the CMB photon number density is $n_{\gamma,\rm BBN}
=20.28\, T_{\rm BBN}^3$ and the energy density is $\epsilon_{\gamma,\rm BBN} =
a\, T_{\rm BBN}^4$, $a=7.5657\times
10^{-15}\,\hbox{erg}\,\hbox{cm}^{-3}\,\hbox{K}^{-4}$ being the black-body
radiation density constant and $T_{\rm BBN}$ the CMB temperature at BBN.

The present-day baryon number density $n_{b,0}$ is related to the baryon
density parameter $\omega_b=h^2 \Omega_b$ by;
\begin{equation}
\omega_b = \frac{\langle m_{\rm b}\rangle \ n_{\rm b,0}}{h^{-2} \rho_{\rm crit,0}} \\
\end{equation}
where $h=H_0/100\,\hbox{km}\,\hbox{s}^{-1}\,\hbox{Mpc}^{-1}$ and $\Omega_b=
\rho_b/\rho_{\rm crit} = 8\pi G \rho_b/3 H_0^2$, $G$ being the gravitational
constant and $\rho_b = \langle m_b\rangle n_b$. The mean mass per baryon is
very well approximated by \citep{Fields2020}  $\langle m_b\rangle = (1+\delta)
m_{\rm H}$ where $\delta=-\left[ 1.744  +  7.119  \left( Y - 0.245 \right)
\right] \times 10^{-3}$ accounts for the reduction of the mass due to helium
binding; $Y$ is the primordial helium mass fraction. Combining the above
relations we get \citep{Fields2020}:
\begin{equation}\label{eq:eta}
\eta_{\rm BBN} = \frac{273.754\times 10^{-10}}{1 - 7.131 \times 10^{-3} \left(Y-0.245
\right)} \, \left(\frac{2.7255\rm K}{T_{\rm BBN,0}}\right)^3 \,\omega_b\, ,
\end{equation}
where $T_{\rm BBN,0}$ is $T_{\rm BBN}$ redshifted to $z=0$. Note that this
expression does not contain the Hubble parameter and therefore is not affected
by the current discrepancy between the value derived by \textit{Planck} CMB
anisotropy measurements \citep{Planck2018parameters} and the local value
derived by \citet{Riess2019}.

The excellent agreement between the values of $\eta$ inferred from BBN and from
CMB anisotropies \citep{Planck2018parameters, Fields2020} implies a tight upper
limit to the amount of energy that could have been released to the CMB after
the nucleosynthesis epoch. \citet{Fields2020} found $\eta_{\rm BBN}=(6.084\pm
0.230)\times 10^{-10}$ and $\eta_{\rm CMB}=(6.090\pm 0.060)\times 10^{-10}$,
allowing for variations of the number of neutrino species ($\eta_{\rm
BBN}=(6.143\pm 0.190)\times 10^{-10}$ and $\eta_{\rm CMB}=(6.104\pm
0.058)\times 10^{-10}$ for 3 neutrino species). The latest \textit{Planck} best
fit value of the baryon density for the TT+TE+EE+lowE+BAO data combination is
$\omega_b=0.02242 \pm 0.00014$ \citep[Table\,2 of][]{Planck2018parameters}.

We define $\delta\eta/\eta=(\eta_{\rm BBN}-\eta_{\rm CMB})/\eta_{\rm CMB}$.
Then, after eq.~(\ref{eq:eta}) and the analogous equation at decoupling, the
fractional difference, $\Delta\epsilon/\epsilon_\gamma$, between the photon
energy density at decoupling ($\epsilon_{\gamma,\rm CMB}$) and at BBN
($\epsilon_{\gamma,\rm BBN}$), can be written as:
\begin{equation}
\frac{\Delta\epsilon}{\epsilon_\gamma} \equiv \frac{\epsilon_{\gamma,\rm CMB}-\epsilon_{\gamma,\rm BBN}}{\epsilon_{\gamma,\rm CMB}} \approx
1-\left(\frac{\eta_{\rm CMB}}{\eta_{\rm BBN}}\right)^{4/3}\simeq \frac{4}{3}\frac{\delta\eta}{\eta},
\end{equation}
where we have  assumed $(\Delta\epsilon/\epsilon_\gamma, \,\delta\eta/\eta) \ll
1$ and used the fact that $\epsilon_\gamma\propto n_\gamma^{4/3} \propto
\eta^{-4/3}$.

$\Delta\epsilon/\epsilon_\gamma$ is consistent with zero with an r.m.s.
uncertainty of
\begin{equation}
\sigma(\Delta\epsilon/\epsilon_\gamma)\simeq\frac{4}{3}\left[(\delta\ln\eta_{\rm BBN})^2+
(\delta\ln\eta_{\rm CMB})^2\right]^{1/2}\simeq 0.039,
\end{equation}
so that the 95\% confidence upper limit to the fractional amount of energy
density that could have been added to the CMB after the BBN is 0.078.

A different approach to derive model independent constraints on electromagnetic
energy releases after primordial nucleosynthesis was adopted by
\citet{Chluba2020}. Their argument goes as follows. Deep in the radiation
dominated era, but after the electron/positron annihilation, the expansion
timescale is controlled by the energy density of photons and neutrinos,
$\rho=\rho_\gamma + \rho_\nu$. A difference, $\delta\rho$, between the value of
$\rho$ at BBN and its standard value, due to a subsequent energy injection,
leads to a non-standard expansion rate, impacting on the primordial production
of light elements.

Such difference can be parameterized by a variation of the effective number of
neutrino species, $N_{\rm eff}$ \citep{SimhaSteigman2008}. For
$\delta\ln(N_{\rm eff})\ll 1$ \citet{Chluba2020} found $|\delta\rho/\rho|\simlt
|0.2417\delta\ln(N_{\rm eff})/(f_\nu - 0.4089)|$, where $f_\nu$ is the fraction
of energy injected going into neutrinos. \textit{Planck} measurements of the
CMB anisotropy power spectra yielded $N_{\rm eff}=2.99\pm 0.17$ ($1\,\sigma$
error). Using the 95\% confidence uncertainty, $\delta\ln(N_{\rm eff})=0.112$,
\citet{Chluba2020} found $\Delta\epsilon/\epsilon_\gamma \simlt
0.11$\footnote{The value $\Delta\epsilon/\epsilon_\gamma \simlt 0.077$ quoted
in that paper is a misprint (Jens Chluba, private communication).} in the case
of a release of pure electromagnetic energy ($f_\nu=0$), and
$\Delta\epsilon/\epsilon_\gamma \simlt 0.25$ for $f_\nu=0.5$, both limits being
at the 95\% confidence level. \citet{Chluba2020} also point out that if $f_\nu
\simeq 0.4089$  the energy release is unconstrained by measurements of $N_{\rm
eff}$.

A more precise calculation of constraints on $\Delta\epsilon/\epsilon_\gamma$
should take into account that they depend on the difference between the
effective number of neutrino species derived from BBN \citep[$2.86\pm
0.15$;][]{Fields2020} and that derived from CMB anisotropies, referring to the
much later decoupling epoch \citep{SimhaSteigman2008}. The error on the $N_{\rm
eff,BBN}-N_{\rm eff,dec}$ is a factor of 1.33 higher than the error on $N_{\rm
eff,dec}$ used by \citet{Chluba2020}. This also implies that the future
strengthening of the constraints on $\Delta\epsilon/\epsilon_\gamma$ depends on
decreasing the uncertainty on $N_{\rm eff}$ at both epochs.

The 95\% confidence limit on $\Delta\epsilon/\epsilon_\gamma$ derived in this
paper is illustrated by the black horizontal line in
Fig.~\ref{fig:constraints}. Also shown, for comparison, are the limits obtained
by \citet{Chluba2020} from the uncertainty on $N_{\rm eff}$, for two values of
$f_\nu$. The lines on the left part of the figure show the constraints implied
by upper limits on $\mu$-type distortions, again from \citet{Chluba2020}.


At $z\simlt 4-5\times 10^6$ constraints from COBE-FIRAS supersede those from
BBN. At higher redshifts, thermalization of additional energy injected into the
CMB smooths out distortions, thus weakening or erasing constraints on the
thermal history of the universe. Even releases of very large amounts of energy,
$\Delta\epsilon/\epsilon_\gamma \sim 1$, occurring at $z\simeq 6.5-7.5\times
10^6$ wouldn't leave imprints detectable by COBE/FIRAS \citep{Fixsen2009} or by
a PIXIE-like experiment \citep{KogutFixsen2020}, orders of magnitude more
sensitive.

\section{Conclusions}\label{sect:conclusions}

The constraint from primordial nucleosynthesis, derived in this paper, implies
that no more than 7.8\% of the present-day CMB energy density could have been
released after primordial nucleosynthesis. Much stronger constraints are set by
COBE/FIRAS measurements of the  CMB spectrum, but these are limited to $z\simlt
5\times10^6$ \citep{Chluba2020}, i.e. to redshifts about two orders of
magnitude lower than the nucleosynthesis redshift.

Comparable constraints on post-BBN energy releases have been derived by
\citet{Chluba2020} from measurements of the effective number of neutrino
species. As pointed out by these authors, such constraints are affected by
uncertainties on the fraction, $f_\nu$, of energy going to neutrinos. They are
weak for $f_\nu \approx 0.3-0.5$ and can be completely avoided for
$f_\nu\approx 0.409$, a value close to the typical amount of energy carried by
neutrinos in TeV scale dark matter annihilations.

Constraints on the contribution to $N_{\rm eff}$ by high energy neutrinos from
dark matter decay in the early universe have been obtained by
\citet{AcharyaKhatri2020}.

Tighter constraints on early energy releases from specific processes have been
inferred from measurements of the light element abundances. Examples are
Hawking evaporation of primordial black holes \citep[e.g.,][]{Keith2020} and
decaying particles \citep[e.g.,][]{Kawasaki2020}. However these constraints are
somewhat model dependent as they rest on the details of hadronic and
electromagnetic interactions involved. On the contrary, the limit presented
here is general and model independent.

Above $z\sim 5\times10^6$ thermalization effects wash out spectral distortions
induced by energy injections into the CMB. Planned next-generation experiments
like PIXIE \citep{Kogut2019, Chluba2019}, PRISM \citep{Andre2014},
PRISTINE\footnote{\url{https://www.ias.u-psud.fr/en/content/pristine}} or the
microwave spectro-polarimetry mission proposed by \citet{Delabrouille2019} will
reach sensitivities orders of magnitude higher. However the redshift range over
which spectral distortions are visible will be only marginally extended.

\section*{Acknowledgements}
We are grateful to the referee and to Jens Chluba for useful comments. Jens
Chluba also provided helpful clarifications on the \citet{Chluba2020} paper. MB
acknowledges support from INAF under PRIN SKA/CTA FORECaST and from the
Ministero degli Affari Esteri della Coo\-pe\-ra\-zione Internazionale -
Direzione Generale per la Promozione del Sistema Paese Progetto di Grande
Rilevanza ZA18GR02.

\section*{Data Availability}

The research described in this letter does not make use of any database. The
data used are those available in the references cited when the data are
introduced. The data generated are given in the text.



\bibliographystyle{mnras}
\bibliography{BBN} 

\begin{thebibliography}{}
\makeatletter
\relax
\def\mn@urlcharsother{\let\do\@makeother \do\$\do\&\do\#\do\^\do\_\do\%\do\~}
\def\mn@doi{\begingroup\mn@urlcharsother \@ifnextchar [ {\mn@doi@}
  {\mn@doi@[]}}
\def\mn@doi@[#1]#2{\def\@tempa{#1}\ifx\@tempa\@empty \href
  {http://dx.doi.org/#2} {doi:#2}\else \href {http://dx.doi.org/#2} {#1}\fi
  \endgroup}
\def\mn@eprint#1#2{\mn@eprint@#1:#2::\@nil}
\def\mn@eprint@arXiv#1{\href {http://arxiv.org/abs/#1} {{\tt arXiv:#1}}}
\def\mn@eprint@dblp#1{\href {http://dblp.uni-trier.de/rec/bibtex/#1.xml}
  {dblp:#1}}
\def\mn@eprint@#1:#2:#3:#4\@nil{\def\@tempa {#1}\def\@tempb {#2}\def\@tempc
  {#3}\ifx \@tempc \@empty \let \@tempc \@tempb \let \@tempb \@tempa \fi \ifx
  \@tempb \@empty \def\@tempb {arXiv}\fi \@ifundefined
  {mn@eprint@\@tempb}{\@tempb:\@tempc}{\expandafter \expandafter \csname
  mn@eprint@\@tempb\endcsname \expandafter{\@tempc}}}

\bibitem[\protect\citeauthoryear{{Acharya} \& {Khatri}}{{Acharya} \&
  {Khatri}}{2020}]{AcharyaKhatri2020}
{Acharya} S.~K.,  {Khatri} R.,  2020, arXiv e-prints, \href
  {https://ui.adsabs.harvard.edu/abs/2020arXiv200706596A} {p. arXiv:2007.06596}

\bibitem[\protect\citeauthoryear{{\b{Alpher}} \& {Herman}}{{\b{Alpher}} \&
  {Herman}}{1949}]{AlpherHerman1949}
{\b{Alpher}} R.~A.,  {Herman} R.~C.,  1949, \mn@doi [Physical Review]
  {10.1103/PhysRev.75.1089}, \href
  {https://ui.adsabs.harvard.edu/abs/1949PhRv...75.1089A} {75, 1089}

\bibitem[\protect\citeauthoryear{{Andr{\'e}} et~al.,}{{Andr{\'e}}
  et~al.}{2014}]{Andre2014}
{Andr{\'e}} P.,  et~al., 2014, \mn@doi [\jcap] {10.1088/1475-7516/2014/02/006},
  \href {https://ui.adsabs.harvard.edu/abs/2014JCAP...02..006A} {2014, 006}

\bibitem[\protect\citeauthoryear{{Burigana}, {Danese}  \& {de
  Zotti}}{{Burigana} et~al.}{1991}]{Burigana1991b}
{Burigana} C.,  {Danese} L.,   {de Zotti} G.,  1991, \mn@doi [\apj]
  {10.1086/170479}, \href {http://adsabs.harvard.edu/abs/1991ApJ...379....1B}
  {379, 1}

\bibitem[\protect\citeauthoryear{{Chluba}}{{Chluba}}{2014}]{Chluba2014a}
{Chluba} J.,  2014, \mn@doi [\mnras] {10.1093/mnras/stu414}, \href
  {http://adsabs.harvard.edu/abs/2014MNRAS.440.2544C} {440, 2544}

\bibitem[\protect\citeauthoryear{{Chluba}}{{Chluba}}{2016}]{Chluba2016}
{Chluba} J.,  2016, \mn@doi [\mnras] {10.1093/mnras/stw945}, \href
  {https://ui.adsabs.harvard.edu/abs/2016MNRAS.460..227C} {460, 227}

\bibitem[\protect\citeauthoryear{{Chluba} \& {Jeong}}{{Chluba} \&
  {Jeong}}{2014}]{ChlubaJeong2014}
{Chluba} J.,  {Jeong} D.,  2014, \mn@doi [\mnras] {10.1093/mnras/stt2327},
  \href {http://adsabs.harvard.edu/abs/2014MNRAS.438.2065C} {438, 2065}

\bibitem[\protect\citeauthoryear{{Chluba}, {Khatri}  \& {Sunyaev}}{{Chluba}
  et~al.}{2012}]{Chluba2012}
{Chluba} J.,  {Khatri} R.,   {Sunyaev} R.~A.,  2012, \mn@doi [\mnras]
  {10.1111/j.1365-2966.2012.21474.x}, \href
  {http://adsabs.harvard.edu/abs/2012MNRAS.425.1129C} {425, 1129}

\bibitem[\protect\citeauthoryear{{Chluba} et~al.,}{{Chluba}
  et~al.}{2019}]{Chluba2019}
{Chluba} J.,  et~al., 2019, arXiv e-prints, \href
  {https://ui.adsabs.harvard.edu/abs/2019arXiv190901593C} {p. arXiv:1909.01593}

\bibitem[\protect\citeauthoryear{{Chluba}, {Ravenni}  \& {Acharya}}{{Chluba}
  et~al.}{2020}]{Chluba2020}
{Chluba} J.,  {Ravenni} A.,   {Acharya} S.~K.,  2020, \mn@doi [\mnras]
  {10.1093/mnras/staa2131}, \href
  {https://ui.adsabs.harvard.edu/abs/2020MNRAS.tmp.2235C} {}

\bibitem[\protect\citeauthoryear{{Cyburt}, {Fields}, {Olive}  \&
  {Yeh}}{{Cyburt} et~al.}{2016}]{Cyburt2016}
{Cyburt} R.~H.,  {Fields} B.~D.,  {Olive} K.~A.,   {Yeh} T.-H.,  2016, \mn@doi
  [Reviews of Modern Physics] {10.1103/RevModPhys.88.015004}, \href
  {https://ui.adsabs.harvard.edu/abs/2016RvMP...88a5004C} {88, 015004}

\bibitem[\protect\citeauthoryear{{Danese} \& {de Zotti}}{{Danese} \& {de
  Zotti}}{1977}]{DaneseDeZotti1977}
{Danese} L.,  {de Zotti} G.,  1977, \mn@doi [Nuovo Cimento Rivista Serie]
  {10.1007/BF02747276}, \href
  {https://ui.adsabs.harvard.edu/abs/1977NCimR...7..277D} {7, 277}

\bibitem[\protect\citeauthoryear{{Danese} \& {de Zotti}}{{Danese} \& {de
  Zotti}}{1982}]{DaneseDeZotti1982}
{Danese} L.,  {de Zotti} G.,  1982, \aap, \href
  {http://adsabs.harvard.edu/abs/1982A%26A...107...39D} {107, 39}

\bibitem[\protect\citeauthoryear{{De Zotti}, {Negrello}, {Castex}, {Lapi}  \&
  {Bonato}}{{De Zotti} et~al.}{2016}]{DeZotti2016}
{De Zotti} G.,  {Negrello} M.,  {Castex} G.,  {Lapi} A.,   {Bonato} M.,  2016,
  \mn@doi [\jcap] {10.1088/1475-7516/2016/03/047}, \href
  {https://ui.adsabs.harvard.edu/abs/2016JCAP...03..047D} {2016, 047}

\bibitem[\protect\citeauthoryear{{Delabrouille} et~al.,}{{Delabrouille}
  et~al.}{2019}]{Delabrouille2019}
{Delabrouille} J.,  et~al., 2019, arXiv e-prints, \href
  {https://ui.adsabs.harvard.edu/abs/2019arXiv190901591D} {p. arXiv:1909.01591}

\bibitem[\protect\citeauthoryear{{Fields}, {Olive}, {Yeh}  \& {Young}}{{Fields}
  et~al.}{2020}]{Fields2020}
{Fields} B.~D.,  {Olive} K.~A.,  {Yeh} T.-H.,   {Young} C.,  2020, \mn@doi
  [\jcap] {10.1088/1475-7516/2020/03/010}, \href
  {https://ui.adsabs.harvard.edu/abs/2020JCAP...03..010F} {2020, 010}

\bibitem[\protect\citeauthoryear{{Fixsen}}{{Fixsen}}{2009}]{Fixsen2009}
{Fixsen} D.~J.,  2009, \mn@doi [\apj] {10.1088/0004-637X/707/2/916}, \href
  {https://ui.adsabs.harvard.edu/abs/2009ApJ...707..916F} {707, 916}

\bibitem[\protect\citeauthoryear{{Gamow}}{{Gamow}}{1948}]{Gamow1948}
{Gamow} G.,  1948, \mn@doi [Physical Review] {10.1103/PhysRev.74.505.2}, \href
  {https://ui.adsabs.harvard.edu/abs/1948PhRv...74..505G} {74, 505}

\bibitem[\protect\citeauthoryear{{Hu} \& {Silk}}{{Hu} \&
  {Silk}}{1993}]{HuSilk1993a}
{Hu} W.,  {Silk} J.,  1993, \mn@doi [Physical Review Letters]
  {10.1103/PhysRevLett.70.2661}, \href
  {http://adsabs.harvard.edu/abs/1993PhRvL..70.2661H} {70, 2661}

\bibitem[\protect\citeauthoryear{{Illarionov} \& {Sunyaev}}{{Illarionov} \&
  {Sunyaev}}{1975}]{IllarionovSunyaev1975}
{Illarionov} A.~F.,  {Sunyaev} R.~A.,  1975, \sovast, \href
  {http://adsabs.harvard.edu/abs/1975SvA....18..691I} {18, 691}

\bibitem[\protect\citeauthoryear{{Kawasaki}, {Kohri}, {Moroi}, {Murai}  \&
  {Murayama}}{{Kawasaki} et~al.}{2020}]{Kawasaki2020}
{Kawasaki} M.,  {Kohri} K.,  {Moroi} T.,  {Murai} K.,   {Murayama} H.,  2020,
  arXiv e-prints, \href {https://ui.adsabs.harvard.edu/abs/2020arXiv200614803K}
  {p. arXiv:2006.14803}

\bibitem[\protect\citeauthoryear{{Keith}, {Hooper}, {Blinov}  \&
  {McDermott}}{{Keith} et~al.}{2020}]{Keith2020}
{Keith} C.,  {Hooper} D.,  {Blinov} N.,   {McDermott} S.~D.,  2020, arXiv
  e-prints, \href {https://ui.adsabs.harvard.edu/abs/2020arXiv200603608K} {p.
  arXiv:2006.03608}

\bibitem[\protect\citeauthoryear{{Khatri} \& {Sunyaev}}{{Khatri} \&
  {Sunyaev}}{2012}]{KhatriSunyaev2012}
{Khatri} R.,  {Sunyaev} R.~A.,  2012, \mn@doi [\jcap]
  {10.1088/1475-7516/2012/06/038}, \href
  {http://adsabs.harvard.edu/abs/2012JCAP...06..038K} {6, 38}

\bibitem[\protect\citeauthoryear{{Kogut} \& {Fixsen}}{{Kogut} \&
  {Fixsen}}{2020}]{KogutFixsen2020}
{Kogut} A.,  {Fixsen} D.~J.,  2020, \mn@doi [\jcap]
  {10.1088/1475-7516/2020/05/041}, \href
  {https://ui.adsabs.harvard.edu/abs/2020JCAP...05..041K} {2020, 041}

\bibitem[\protect\citeauthoryear{{Kogut}, {Abitbol}, {Chluba}, {Delabrouille},
  {Fixsen}, {Hill}, {Patil}  \& {Rotti}}{{Kogut} et~al.}{2019}]{Kogut2019}
{Kogut} A.,  {Abitbol} M.~H.,  {Chluba} J.,  {Delabrouille} J.,  {Fixsen} D.,
  {Hill} J.~C.,  {Patil} S.~P.,   {Rotti} A.,  2019, in Bulletin of the
  American Astronomical Society. p.~113 (\mn@eprint {arXiv} {1907.13195})

\bibitem[\protect\citeauthoryear{{Planck Collaboration VI}}{{Planck
  Collaboration VI}}{2020}]{Planck2018parameters}
{Planck Collaboration VI} 2020, \mn@doi [\aap] {10.1051/0004-6361/201833910},
  \href {https://ui.adsabs.harvard.edu/abs/2020A&A...641A...6P} {641, A6}

\bibitem[\protect\citeauthoryear{{Riess}, {Casertano}, {Yuan}, {Macri}  \&
  {Scolnic}}{{Riess} et~al.}{2019}]{Riess2019}
{Riess} A.~G.,  {Casertano} S.,  {Yuan} W.,  {Macri} L.~M.,   {Scolnic} D.,
  2019, \mn@doi [\apj] {10.3847/1538-4357/ab1422}, \href
  {https://ui.adsabs.harvard.edu/abs/2019ApJ...876...85R} {876, 85}

\bibitem[\protect\citeauthoryear{{Simha} \& {Steigman}}{{Simha} \&
  {Steigman}}{2008}]{SimhaSteigman2008}
{Simha} V.,  {Steigman} G.,  2008, \mn@doi [\jcap]
  {10.1088/1475-7516/2008/06/016}, \href
  {https://ui.adsabs.harvard.edu/abs/2008JCAP...06..016S} {2008, 016}

\bibitem[\protect\citeauthoryear{{Steigman}}{{Steigman}}{2007}]{Steigman2007}
{Steigman} G.,  2007, \mn@doi [Annual Review of Nuclear and Particle Science]
  {10.1146/annurev.nucl.56.080805.140437}, \href
  {https://ui.adsabs.harvard.edu/abs/2007ARNPS..57..463S} {57, 463}

\bibitem[\protect\citeauthoryear{{Sunyaev} \& {Zeldovich}}{{Sunyaev} \&
  {Zeldovich}}{1970}]{SunyaevZeldovich1970}
{Sunyaev} R.~A.,  {Zeldovich} Y.~B.,  1970, \mn@doi [\apss]
  {10.1007/BF00653472}, \href
  {http://adsabs.harvard.edu/abs/1970Ap%26SS...7...20S} {7, 20}

\bibitem[\protect\citeauthoryear{{Tashiro}}{{Tashiro}}{2014}]{Tashiro2014}
{Tashiro} H.,  2014, \mn@doi [Progress of Theoretical and Experimental Physics]
  {10.1093/ptep/ptu066}, \href
  {https://ui.adsabs.harvard.edu/abs/2014PTEP.2014fB107T} {2014, 06B107}

\bibitem[\protect\citeauthoryear{{Zeldovich} \& {Sunyaev}}{{Zeldovich} \&
  {Sunyaev}}{1969}]{ZeldovichSunyaev1969}
{Zeldovich} Y.~B.,  {Sunyaev} R.~A.,  1969, \mn@doi [\apss]
  {10.1007/BF00661821}, \href
  {http://adsabs.harvard.edu/abs/1969Ap%26SS...4..301Z} {4, 301}

\makeatother
\end{thebibliography}




\bsp	
\label{lastpage}
\end{document}